\documentclass[russian,reqno,dvips,12pt]{article}

\usepackage{epsfig,graphicx,amsmath,amsfonts,amssymb,amsthm}

\theoremstyle{plain}

\theoremstyle{remark}
\newtheorem{remark}{\qquad Remark}
\newtheorem{example}{\qquad Example}
\newtheorem{definition}{\qquad Definition}

\begin{document}

\title{Unbounded Probability Theory and Its Applications}

\author{V.~P.~Maslov and T.~V.~Maslova\\
{\small Moscow Institute for Electronics and Mathematics at HSE}}

\date{}

\maketitle

\begin{abstract}
The paper deals with the order statistics and empirical mathematical expectation
(which is also called the estimate of mathematical expectation in the literature)
in the case of infinitely increasing random variables.
The Kolmogorov concept which he used in the theory of complexity
and the relationship with thermodynamics which was pointed out already
by Poincar\'e are considered.

The mathematical expectation
(generalizing the notion of arithmetical mean,
which is generally equal to infinity for any increasing sequence of random variables)
is compared with the notion of temperature in thermodynamics
by using an analog of nonstandard analysis.

The relationship with the Van-der-Waals law of corresponding states is shown.
Some applications of this concept in economics,
in internet information network, and self-teaching systems
are considered.

\bigskip
Key words: theory of complexity, thermodynamics, Bose gas, Wiener quantization,
Maxwell rule, nonstandard analysis, law of corresponding states,
distribution according to income, Monte Carlo method, queuing theory, World Wide Web.
\end{abstract}

\bigskip

\section{Introduction}

Let us recall the history of unrestricted probability theory and
quote, in greater length than in previous publications, some
writings of the great mathematician H.Poincar\'e, who was, at the
time of these writings, the chairman of the department of
probability theory and mathematical statistics at the Sorbonne. In
his book {\it Science et Hypoth\`ese} in Chapter XI, entitled
{\it Calcul des Probabilit\'es}, Poincar\'e writes:

``Was probability defined? And can it be defined? And if not, how can
we dare argue about it? The definition, one may say, is very simple:
the probability of an event is the ratio of the number of favorable
outcomes to the total number of outcomes.

A simple example shows how incomplete this definition is. Suppose I
throw two dice; what is the probability that at least one of them
will come up with 6? Each dice can come up with six different
values: the total number of outcomes is $6\times6=36$; the number of
favorable ones is $11$, thus the probability is $11/36$.

Such is the correct solution. But I will be equally correct if I say
that the throw of the two dice can result in $(6\times7)/2=21$
different outcomes; among these $6$ are favorable; thus the
probability is $6/21$.\footnote{Note that in this (second) calculation
the outcome (1,6) is identified with the outcome (6,1), i.e.,
it is counted only once; in other words, {\it we do not distinguish the two dice}.
We shall see below that, in thermodynamics,
this corresponds from Boltzman statistics to the statistics of Bose-Einstein.}

Why is the first method of calculating the probability better than
the second one? Whatever the case, our definition doesn't tell us.

Thus we must complete the definition by saying ``\dots to the total
number of outcomes provided these outcomes are equiprobable". And so
we come to a definition of probability in terms of probability itself.

How can we find out whether or not two events are equiprobable?
Isn't that the result of some arbitrary convention? If, at the
beginning of each problem, we specify what the arbitrary convention
is, then all will work out well, we will only need to apply the
rules of arithmetic and algebra and perform the corresponding
computation, whose result will then leave no room for any doubt. But
if we wish to apply this result, we will have to prove that our
arbitrary convention was valid, and so we return to the difficulty
that we had tried to circumvent.

One might say: a bit of common sense will suffice to decide what
specific convention should be chosen. However, Bertrand, out of pure
curiosity, studied the following simple problem: ``What is the
probability that a chord of the circle is longer than the side of an
inscribed equilateral triangle?" The great geometer made two
conventions, both suggested by common sense, and obtained $1/2$ in
one case and $1/3$ in the other.

The conclusion that apparently follows from the above is that
probability theory is a useless science, that we should not trust
the unclear instinct that we call common sense and to which we
appeal in order to specify our conventions"
(English translation of~\cite{Poincare}, pp.~115--116).

The difficulties related to the dependence of probability theory on
common sense, which were already pointed out by Laplace, were
overcome by Kolmogorov in his complexity theory. We toss a coin but,
instead of of talking about the probability of getting heads or
tails, we perform a sufficiently large number of long trials and
then notice that most of the possible combinations lie in the
interval $1/2\pm\sqrt N$, where $N$ is the number of trials in the
sequences. Here Kolmogorov finds the shortest code that determines
an individual sequence of trials, and the length of that code will
be close to the value of the entropy.

This means, although Kolmogorov does not say it outright, that his
approach is based on entropy.

This remarkable achievement, in the case when the values of the
random variable are not unbounded, completely answers the question
asked by Poincar\'e in his philosophical work quoted above.

Here we carry over Kolmogorov's approach to sequences of random
variables in the case when the sequences infinitely increase in
length.

First we consider the case of infinitely growing values of the
random variable corresponding to income, i.e., the case of the
so-called ``Petersburg paradox". A player plays on a roulette; he can
put his stake on black or red, and begins by putting $l$ dollars on
black. If the roulette stops on red, he plays black again,
increasing his stake to $el$. Having lost, he continues playing
black, again multiplying his stake by $e$, and continues in this way
until, after $m$ steps, the roulette stops on black. The stake at
that point was $e^ml$. The player lost
$$
\sum_{k=0}^{m-1} e^k l= l\frac{e^m-1}{e-1},
$$
and won
$$
e^ml.
$$
His final winnings were therefore
$$
0.418e^m \cdot l, \qquad  m>2.
$$
In its time this paradox could not be used by roulette players to
obtain garanteed winnings, because they did not have sufficient sums
of money. But today, when money is electronic and high level
functionaries in Russia have unlimited credit in state banks, the
possibility of enrichment in such ways is boundless
\footnote{ In France, Germany and other countries there are always
ongoing cases in courts of law where such wrongdoings by high level
functionaries are judged. }. And if society does not have sufficient
control over the actions of its functionaries, this is very
dangerous for the economy.

If winnings can be regarded as a random variable, more precisely, as
the values of some random variable $\xi$, then the exponential
growth of such a random variable is catastrophic.

In our times, we are nearer to a dangerous economic catastrophe than
to the ecological one everybody is talking about. In 1567, the
(saint) Pope Pius~V wrote the bull ``De salutis gregio dominicis''
(On the salvation of the Lord's flock). But neither the inquisition,
nor the laws of Shiriat, are capable of saving all of society, which
consists of people of different mentalities and different religious
beliefs (see Remark~\ref{rem6} below).

We would like to point out that historians do not pay sufficient
attention to such a problem as corruption. Marxist historians
regarded ethical wrongdoings and the corruption of high level
functionaries as too rough an explanation of social problems.

But already Salustius (86--35\,BC) believed that the principal reason for
the fall of the first oligarchy was the all engulfing corruption. As
an example of the implementation of strict Roman ethical rules he
indicates Caton (on the role of ethical rules in society, see~\cite{Gusein}).

\begin{remark}\label{rem1}
In dealing with the order statistics of the income of one of the member
countries of the European Union, one should consult the statistical
data of the country before it became a member of the EU. Then one
can predict, by using unrestricted probability theory, when this
country will ``fall out into a condensate" like dew condensing from
the mixture of gases constituting air.
\end{remark}

\section{Dependence on the choice of the origin.
Notion of temperature as mean energy}

The great H.Poincar\'e pointed out the relationship between
thermodynamics and temperature. He wrote: ``Let me also mention the
kinetic theory of gases and the well known hypothesis, according to
which it is assumed that each molecule of gas describes a very
complicated trajectory, whereas the phenomena considered in the mean
(the only way that they can be observed) are subject, in view of the
law of large numbers, to simple rules, like the laws of Mariotte and
Gay-Lussac.

All these theories are founded on the law of large
numbers, and so the failure of probability theory would lead to that
of these laws." (English translation of \cite{Poincare}, p.~117)\footnote{Poincar\'e
constructed a mathematical model of ideal gas.
V.V.Kozlov in~\cite{Kozlov_2009} proved that the Gibbs paradox
takes place for this model.
Kozlov's proof stimulated investigations of this important problem by ]
one of the authors (see~\cite{MZ_83-5},~\cite{RJ-17-3},~\cite{MZ_89_3}, etc.).}.

In our work, we construct a probability theory which is intimately
related with the kinetic theory of gases and is not based on the
notion of equiprobability of events. It follows Kolmogorov's main
idea, which he formulated in the framework of his complexity theory.

Let us clarify the relationship with Bose gas.

The empirical expectation of a random variable taking the values
$\{\lambda_i\}$ has the form
\begin{equation}\label{1j}
\sum_{i=1}^\infty \frac{n_i}{n}\lambda_i = \mathcal{E}, \qquad
\sum_{i=1}^\infty \frac{n_i}{n}=1,
\end{equation}
or equivalently
\begin{equation}\label{2j}
\sum_{i=1}^\infty {n_i}\lambda_i =M, \qquad M=n \mathcal{E},
\end{equation}
\begin{equation}\label{3j}
\sum_{i=1}^\infty {n_i}=n.
\end{equation}
These equations coincide precisely with the equations used to
determine the inner energy of an ideal gas if $\lambda_i$ is the
collection of eigenvalues of a certain unbounded self-adjoint
operator $\widehat H$, known as the Hamiltonian,
and $N_i$ is the number
of particles at the corresponding level:
\begin{equation}\label{4j}
\sum_{i=1}^\infty {N_i}\lambda_i =E, \qquad
\sum_{i=1}^\infty {N_i}=N.
\end{equation}

The mean energy of one particle is
$$
E_{\text{average}} =\frac{E}{N}.
$$
As is well known, it is related with the notion of
{\it temperature}~$T$.

Indeed, we all know that in a gas the number of particles is huge:
in a vessel of volume $V=1\text{cm}^3$ there are $N\sim10^{19}$
particles (at room temperature and under pressure of $1$\,atm).
The mean energy of the gas per unit volume is $kT$,
where $k$ is the Boltzman constant and $T$ is the temperature of gas in its
equilibrium state.

Therefore, in thermodynamics, a mean quantity corresponding to a
huge number of tests has been defined in the case when the random
variable $\lambda$ assumes unboundedly growing values. Its growth
will be discussed below.

Van der Vaals is famous for his empirical relation (the VdV equation).
But all surveys articles of thermodynamics claim that van
der Vaals' main discovery is the law of corresponding states.
This law consists in the following. Real pure\footnote{ A pure gas is a gas consisting of identical molecules. }\;
gases have the same temperature $T_c$ and the same critical pressure~$P_c$.

D.Mendeleev referred to the critical temperature $T_c$ as the
temperature of absolute boiling. He had in mind that the liquid
state is impossible at higher temperatures. What is temperature (in
the Kelvin scale)? It is counted off from absolute zero. However,
the absolute zero is unaccessible in the sense that no matter how
small the temperature (in the Kelvin scale), it still is infinitely
far from absolute zero. In this sense it is more natural to consider
the logarithm of the temperature. Then $\log T|_{T=0}=-\infty$,
and it is obviously seen that the distance between the temperature
(in the Kelvin scale), no matter how small,
and absolute zero is an infinite quantity.
Thus we are dealing, from the mathematical standpoint,
with ``nonstandard analysis" and an infinite temperature, although
the latter may equal, say, one hundredth of a degree (in the Kelvin scale).

If we pass to the relative quantities
\begin{equation}\label{4ja}
T_{red}= \frac{T}{T_c}, \qquad  P_{red}= \frac{P}{P_c},
\end{equation}
then it turns out that in these reduced coordinates,
the isotherms, isochores, and isobars on all the
coordinate planes (pressure -- volume), (temperature -- density) and
so on behave very similarly for different gases even when the gases
have very different critical temperatures $T_c$. This is true for
nitrogen, argon, carbon dioxide, methane and other gases.

\begin{example}\label{ex1}
Consider the standard Hilbert space $H$ of functions defined in the
closed interval $[0,\pi]$, and its basis of orthogonal functions
\begin{equation}\label{5j}
\{ e^{inx}\}^{n=\infty}_{n= - \infty},
\end{equation}
and the following two self-adjoint operators
\begin{equation}\label{6j}
\widehat{A}_1=-ih \frac{\partial}{\partial x}, \qquad
\widehat{A}_2= -2 ih \frac{\partial}{\partial x},
\end{equation}
depending on the parameter $h=1/k$, where $k$ is an integer.

The initial reference point in the basis $\{e^{inx}\}$ is $n= 1$.

Consider $n_0^{(1)}=k$ and $n_0^{(2)}=\frac{k}{2}$. We see that
$$
\widehat{A}_1 e^{in_0^{(1)}x}\approx 1,
$$
$$
\widehat{A}_2 e^{in_0^{(2)}x}\approx 1.
$$
Obviously,
$$
\big\{ e^{i(n-k)x}\big\}^{n=\infty}_{n=-\infty}
$$
and
$$
\big\{ e^{i(n-k/2)x}\big\}^{n=\infty}_{n=-\infty}
$$
are isomorphic to
$$
\big\{ e^{inx}\big\}^{n=\infty}_{n=-\infty}.
$$

Thus, as $h\to0$, the reference point of the infinite basis shifts
from the point $n=1$ to the point $n=1/h$ in the first case
$\widehat A_1$ and to the point $n=1/(2h)$ in the second case
$\widehat A_2$. On the interval $[0,\pi]$, the operators
$\widehat A_1$ and $\widehat A_2$ will act exactly in the same way
on functions $x$ expressed in the basis~\eqref{5j} as the operator
$\widehat A$ of the usual numbering of functions. When we shift the
numbering by $n-k$ and $n-k/2$, we come to rapidly oscillating
functions. These two spaces are orthogonal to each other and to the
initial Hilbert space $H$ (in the sense that
$\int\sin\frac xh\varphi(x)\,dx\to0$ as $h\to0$, $\psi(x) \in H$).

Denoting $e^{ikx} =\Psi_c$ and $e^{ik/2x} =\varphi_c$,
we see that the space of functions corresponding to the basis of the form
$\Psi_n/\Psi_c$ is isomorphic to the basis of functions
$\Psi_n/\varphi_c$ as $h\to0$, where $\Psi_n= e^{inx}$
($h=0$, $k=\infty$  in the ``nonstandard analysis'').

A similar situation occurs for Dirichlet series and for the
eigenfunctions of the Schr\"odinger operator (the Hamilton operator)
$\widehat H$) for the oscillator and so on\footnote{
Note that the relationship between the solution of the
Cauchy problem for the Schr\"odinger equation of the form
$e^{i\widehat Ht}\Psi_0$, where $\Psi_0\in H$ is the initial
condition, and the solution of the heat equation
$e^{-\widehat Ht}\Psi_0$, is described in Section~2 of the paper~\cite{TMF_170-3}
which deals with Wiener quantization. }\; (see~\cite{DAN_1957}).

This isomorphism of infinitely large quantities from ``nonstandard
analysis" explains, in a certain sense, the empirical law of
corresponding states due to Van der Vaals, who discovered a similar
``almost isomorphism".

It is difficult for a mathematician to explain the ideas of the new
thermodynamics to a physicist. The most difficult thing to explain
is that such a small temperature as $T=1/100\,K$ is actually an
infinitely large quantity. For the expectation~\eqref{1j}
of a random variable taking unboundedly increasing values
(as, for instance, in the case $\lambda_i=i^2$, $i=1,2,\dots$),
this fact is quite natural
in the case when we decompose the natural number $M$ into $N$
summands, as well as in the case of frequency dictionaries~\cite{NTI}.
\end{example}

\begin{example}\label{ex2}
Let us recall a famous theorem from number theory, namely the
solution of the ``partitio numerorum'' problem. In this problem we
deal with a natural number $M$ that we are to decompose into the sum
of $N$ summands (positive integers), for example $M=5$, $N=2$:
$$
 5=1+4=2+3,
$$
which gives 2 variants of the decomposition: $\mathcal{M}=2$.

If $M=10^{23}$, $N=1$, there is only one variant: $\mathcal{M}=1$.
If $M=10^{23}$ and $N=10^{23}$,
there is also only one variant of the decomposition (the sum of ones): $\mathcal{M}=1$.

Obviously, for a fixed $M$ there exists a number $N_c$ for which the
number of variants $\mathcal M$ is maximal (this number is not unique in
general). The quantity $\log_2 \mathcal{M}$ is called {\it Hartley entropy}.
The point at which it achieves its maximum is where the
maximal entropy occurs.

Suppose we are given the decomposition
$$
M= a_1+ \cdots + a_N
$$
of the number $M$ into $N$ summands. Denote by $N_j$ the number of
summands in the right-hand side of that equation which are exactly
equal to~$j$.

Then there will be $\sum_jN_j$ summands, and this sum equals $N$,
since there are $N$ summands in all. Further, the sum of all
summands that equal $j$ is $jN_j$, since there are $jN_j$ summands
$N_j$, so that the sum of all the summands is obtained by summing
all these expressions over $j$, i.e., it is $\sum_jjN_j$, which
equals $M$. Namely
\begin{equation}\label{1x}
\sum_{i=1}^\infty  N_i=N, \qquad  \sum_{i=1}^\infty  i N_i=M.
\end{equation}

The nonuniqueness of the above mentioned maximum and the
indeterminacy of their quantity permitted Erd\"os to obtain his
result only with precision $o(\sqrt{M})$.
\end{example}

Thus we have the Erd\"os theorem for the system of two Diophantine equations
\begin{equation}\label{df-eq}
\sum_{i=1}^\infty N_i=N, \qquad  \sum_{i=1}^\infty i N_i=M,
\end{equation}
which asserts that the maximum number of solutions $N_c$ of this
system is
\begin{equation}\label{8x}
N_c=\beta^{-1}M_c^{1/2} \log\, M_c+ \alpha M_c^{1/2}+o(M_c^{1/2}),
\qquad \beta=\pi\sqrt{2/3},
\end{equation}
where the coefficient $\alpha$ is determined by the formula
$\beta/2= e^{-\alpha\beta/2}$.

If we increase the number $N$ in problem~\eqref{df-eq}
while $M$ remains constant, then the number of solutions will decrease.
If we count the sums~\eqref{df-eq} from one instead of from zero, i.e.,
\begin{equation}\label{6}
\sum_{i=0}^\infty  i N_i = (M-N), \qquad  \sum_{i=0}^\infty  {N_i} = N,
\end{equation}
then the number of solutions will not decrease, and will remain
constant.

Let us explain this fact. The Erd\"os--Lehner problem~\cite{Erdos-Leh}
consists in decomposing the number $M_c$ into $N\le N_c$ summands.

The decomposition of the number 5 into 2 summands gives two
variants. If we include the number 0, we obtain three variants:
$5+0=3+2=4+1$. Thus the inclusion of zero gives us the possibility
of saying we are decomposing $M$ into $k\le n$ summands. Indeed, the
decomposition of 5 into three summands includes all the previous
variants: 5+0+0, 3+2+0, 4+1+0 and adds new variants without zero.

The maximum does not change by much in this case~\cite{Erdos-Leh}, whereas
the number of variants does decrease: the zeros will allow the
maximum to remain constant, and the entropy will not decrease, it
will become constant after the maximum is reached. This remarkable
property of the entropy is what allows us to construct the
unrestricted probability theory in the general case~\cite{MN_91-5}.
In physics, this effect is identical to the so-called Bose condensate
effect.

One may ask: How does the ``partitio numerorum'' in arithmetic
differ from the Boltzman-- Shannon statistics? If we regard $4+1$
and $1+4$ as different variants, then we obtain the
Boltzman--Shannon statistics. The number of variants $\mathcal M$ of the
decomposition rapidly increases. Thus the ``noncommutativity" of
addition gives a huge number of extra variants of the decomposition,
and the Hartley entropy, equal to the logarithm of the number of
variants, will coincide with the Boltzman--Shannon entropy.

The entropy arising in arithmetic, because of commutativity and
symmetry, is the Bose--Einstein entropy.
Let us recall the formula for the entropy of a nonequilibrium Bose
gas, derived in the \emph{textbook}~\cite{Landau_St-ph} (formula~54.6):
\begin{equation}\label{L-L-1}
S=\sum_j G_j [(1+\bar{n}_j) \ln (1+\bar{n}_j)-\bar{n}_j \ln \bar{n}_j],
\end{equation}
where $\bar{n}_j$ are the average numbers of fillings of the quantum
states $n_j=N_j/G_j$.

We have
\begin{gather}
\sum_j N_j =\sum_j G_j \bar{n}_j=N,  \nonumber \\
\sum_j \varepsilon_j N_j =\sum_j \varepsilon_j G_j \bar{n}_j=E. \label{7j}
\end{gather}
From this, applying the method of Lagrange multipliers to the
maximum of the entropy~\eqref{L-L-1} under conditions~\eqref{7j},
we obtain\footnote{We repeat the derivation word for word~\cite{Landau_St-ph}.
A rigorous proof of fractional dimensions was first obtained by
A.M.Vershik~\cite{Vershik}. Also see~\cite{Masl_Naz_83-2}.}
\begin{equation}\label{8j}
\bar{n}_j= \frac{1}{e^{a+ b\varepsilon_j}-1}.
\end{equation}
Here $a= - \mu/T$, $b= 1/T$, $T$ is the temperature,
$\mu \leq 0 $ is the chemical potential.
Thus the Lagrange multipliers are expressed in terms
of the temperature and the chemical potential of the gas.

Since, according to the Bose--Einstein formulas sums of the form
$$
\sum_{i=0}^\infty   \lambda_i
n_i = M, \qquad  \sum_{i=0}^\infty  {n_i} = n,
$$
are replaced by integrals, it is impossible to distinguish the point
$i=0$ itself, the condensation effect must occur in a neighborhood
of this point, not at the point $i=0$ itself.
But for $D \leq 2$,
the Euler--Maclaurin formula cannot be applied (and so we cannot
pass from sums to integrals), and therefore the point $i=0$ can
indeed be distinguished.

However, will the extra ``particles" condense precisely near the
point $i=0$ for $N>N_c$? Actually, this question consists in the
following: will the number of zeros in the problem under
consideration as $M\to\infty$ be much larger than the number of
ones, i.e., will we have
$$
N_1= o(N_0)
$$
even for $N>N_c$?

A computer calculation gives a negative answer. Experts in number
theory also answer negatively. All the more when $D>2$ and the
Euler--Maclaurin formula can be applied, for $i\gg1$ we can pass to
the intergral and it is impossible to speak of one point $i=0$, as
it is done in the textbook~\cite{Landau_St-ph}.

Actually, together with the convergence to the integral (for
instance, in the 3-dimensional case) as
$N\to\infty$ and $\mu\to 0$,
the bell-shaped function $\sqrt{N}e^{-Nx^2}$ also increases, where
\begin{equation}\label{9j}
\frac{1}{\ln N} > x\geq 0
\end{equation}
(cf.~\cite{RJ_19-3-VM}).

In this general argument it is important to understand the meaning
of $G_j$ in the general case.

In the mathematical literature, there are two definitions of $G_j$:
the statistical one, corresponding to quantum statistics, and the
one from mathematical physics, corresponding to the $D$-dimensional
Laplace equation in a $D$-dimensional volume. In exactly the same
way, the Landau--Lifshits textbook has two definitions of the Bose
gas. One appears in the part called ``Statistical Physics", the
other one (which uses the symmetric eigenfunctions of the
Schr\"odinger operator) appears in the part entitled ``Quantum
Mechanics". In the mathematical literature, the first is named after
Weyl, the second, after Courant. Let us look at both definitions in
more detail.

The $2D$-dimensional phase space is discretized into a lattice and
$G_j$ is determined from the formula
\begin{equation}\label{l1}
G_i=\frac{\Delta p_j \Delta q_j}{(2\pi \hbar)^D}.
\end{equation}
Here $\hbar$ is the Planck constant.

By assigning to $G_i$ the multiplicities of the spectrum of the
Schr\"odinger operator, we obtain a correspondence between the
symmetric (with respect to permutations) eigenfunctions of the\linebreak
$N$-particle Schr\"odinger equation with the combinatorial
calculations of the Bose statistics appearing
in the book~\cite{Landau_St-ph}.

The single-particle $\psi$-function satisfies the free Schr\"odinger
equation with Dirichlet conditions on the walls of the vessel.
According to the classical Courant formula,
the number of eigenvalues less than a given $\lambda\gg1$
(the spectral density $\rho(\lambda)$) has the following asymptotics:
\begin{equation}\label{cour-2}
\rho(\lambda)= \frac{Vm^{D/2}\lambda^{D/2}}{\Gamma(D/2+1)(2\pi)^{D/2} \hbar^D} (1+o(1))
\qquad \text{as} \quad \lambda\to\infty
\end{equation}
($V$ is the $D$-dimensional volume).

This is precisely by means of this correspondence that we establish
the relationship between the Bose--Einstein combinatorics~\cite{Landau_St-ph}
with the definition of the $N$-particle Schr\"odinger equation and
the multiplicity of the single-particle Schr\"odinger equation.

The spectrum of the single-particle Schr\"odinger equation without
the interaction potential coincides up to a factor with the spectrum
of the Laplace operator.

In the case of the Laplace operator and the Schr\"odinger equation,
it makes no sense to consider fractional dimensions. In the
statistical case, fractional dimensions do make sense, but their
meaning is different. They determine the average number of degrees
of freedom, which can be different for molecules of different
velocities, so that the average value of the degrees of freedom for
the whole gas can be fractional. It corresponds to the mean energy,
which is the temperature. For the unrestricted probability theory
that we consider here, it is precisely the fractional ``dimension"
(understood as the average of the number of degrees of freedom over
the entire ``general" population) that makes sense.

One of the authors proved that both the degeneration energy for the Bose gas
and the number of particles in the degeneration state coincide up to normalization
with the critical values $N_c$ and $E_c$ (i.e., for $\mu=0$)
of the classical noninteracting gas.
In the $D$-dimensional case, the degeneration energy has the form
\begin{equation}\label{cour-3}
E_d= \int_0^\infty \frac{\frac{|p|^2}{2m}\,
d\varepsilon}{e^{\frac{|p|^2}{2m}/T_d}-1},
\end{equation}
where
\begin{equation}\label{cour-4}
d\varepsilon=\frac{|p|^2}{2m} \frac{dp_1 \dots dp_D \, dV_D}{(2\pi \hbar)^D},
\end{equation}
which implies the coefficients $C$ and $\Lambda^D$ in the formula
\begin{equation}\label{cour-5}
E_d =C\Lambda^D T_d^{2+\gamma}\zeta(1+D/2)\Gamma(1+D/2).
\end{equation}
It is more convenient to pass to another normalization
\begin{equation}\label{nor1}
d\varepsilon'=\frac{1}{\Gamma(D/2+1)}\,d\varepsilon
\end{equation}
for the energy and, respectively,
\begin{equation}\label{nor2}
d\varepsilon''=\frac{1}{\Gamma(D/2)}\,d\varepsilon
\end{equation}
for the number of particles~$N$.

Passing to the Van-der-Waals coordinate~\eqref{4ja},
we obtain the coefficient $C\Lambda^D$
of the form~$\widetilde{\Lambda}^{\gamma-\gamma_c}$,
where $\gamma_c$ is determined from the experimental values of the quantity
$Z|_{\mu=0}=\zeta(\gamma_c+2)/\zeta(\gamma_c+1)$
and $\zeta(x)$ is the Riemann zeta function.
In what follows, we omit the tilde on the coefficient $\Lambda$
and the index $r$ (reduced) on the parameters~$T$ and~$P$.

After this, it is more convenient to use the polylogarithm
$\operatorname{Li}_{2+\gamma} (e^{-\mu/T})$ for the energy
and the polylogarithm~$\operatorname{Li}_{1+\gamma} (e^{-\mu/T})$
for the number of particles:
\begin{gather}\label{nor3}
E= T^{2+\gamma} \operatorname{Li}_{2+\gamma}(a), \\
N= T^{1+\gamma} \operatorname{Li}_{1+\gamma}(a),
\end{gather}
where $a$ is the activity ($a=e^{-\mu/T}$).

The the compressibility factor is equal to
\begin{equation}\label{nor4}
Z=\frac{E}{NT}
=\frac{\operatorname{Li}_{2+\gamma}(a)}{\operatorname{Li}_{1+\gamma}(a)}
=\frac{M}{Tn}.
\end{equation}
The coefficient $\Lambda$ is here eliminated.

\section{Basic series and phase transition}

To each degree of freedom we assign its basis sequence,
whose terms correspond in thermodynamics to pure gases with identical molecules.
Namely, we consider the following basis family of random
variables $\{i^{D/2}\}$, corresponding to the number $D$ of degrees of freedom.

The variational series
\begin{equation}\label{7}
\xi_i=i^{D/2},
\end{equation}
will be called the {\it basis series}\footnote{Similarly to thermodynamics,
the pure gases correspond to the periodic table,
and each pure gas is associated with a certain value~$Z_c$,
for example, up to~$0.01$\%.}
of dimension~$D$.

Let $n_i$ be the value of the outcome corresponding to~$\xi_i$.

A sample value (estimate) of the mathematical expectation is
$$
\frac{\sum n_i \xi_i}{n}=\mathcal{E}.
$$
We assume that $\mathcal{E}\to\infty$ as $n\to\infty$.

It is useful to introduce the parameter $\gamma$ as follows
$$
D=2(\gamma+1), \qquad \gamma= \frac D2 -1.
$$
We shall consider the one-parameter family of entropies $S_\gamma$
as the logarithm of the number of solutions of the system
\begin{equation}\label{8}
\sum_{i=1}^\infty n_i=n, \qquad     \sum_{i=1}^\infty n_i\xi_i\leq n\mathcal{E}.
\end{equation}

The asymptotic relation $n(\mathcal{E})$ between $n$ and $\mathcal{E}$
as $n\to\infty$ for which the maximal number of solutions of~\eqref{8}
(up to $o(N)$) will be called the {\it point of maximum desaccumulation}
of the estimate (sample value) of the expectation,
while the families of solutions $\{n_i\}/n$ of equation~\eqref{8}
will be called {\it typical}; the value $n$ for which the
maximal deacccumulation is achieved is denoted by~$N_{\mathrm{c}}$.

The binary logarithm of the number of solutions of equation~\eqref{8}
is the Hartley entropy~$S$.

Thus $a$, $b$ are the Lagrange multipliers
(see~\eqref{L-L-1}--\eqref{8j}) for finding the maximum of $S_\gamma$
under conditions~\eqref{8}:
\begin{equation}\label{9}
 dS= a\, dn+ b\, d(\mathcal{E}n), \qquad  \mathcal{E}n=M,
\end{equation}
\begin{equation}\label{10}
d(\mathcal{E}n)= \frac{1}{b}\, dS - \frac{a}{b}\, dn.
\end{equation}
Denote the quantity $1/b$ by $T$, and the quantity~$a/b$, by ${\mu}$
($\mu\leq 0$). In thermodynamics, there is an additional term $P\,dV$,
where $P$ is the pressure and~$V$ is the volume.

\begin{remark}\label{rem3}
In the economic calculation of income (see below), the variable $T_c$ will be
called the {\it minimal wage}, while $\bar{\mu}=-\mu$
will be the nominal percentage or the desaccumulation coefficient.
We also introduce the notation $b=1/T$.
To the value $n=N_{\mathrm{c}}$ corresponds $T=1$. For a given $E$,
the maximum of desaccumulation is attained at the value of~$n(E)$,
which we denote by~$N_c$.
To the volume~$V$ corresponds to volume of goods $Q$,
and to the pressure $P$, their price (Irving Fisher's correspondence law).
\end{remark}

The Hartley entropy $S_\gamma$ for $\gamma>0$ has the form
\begin{equation}\label{11}
S=n\bigg[Z(\gamma)(2+\gamma) + \frac{{\mu}}{T}\bigg].
\end{equation}
For ${{\mu}}= 0$ and for $n > N_{\mathrm{c}}$,
the entropy $S$ can be extended continuously by a constant
(a phenomenon that, in the particular case $D=3$,
is equivalent to the physical notion of ``Bose--Einstein condensate").

The relation $n>n_c$ is equivalent to $T< 1$. Here for $n \gg k \gg 1$
(see~\eqref{9j}), the number of values greater than the critical one is
$\sum_{i=k}^\infty n_i= n_c$. The remaining $n-n_c$
values accumulate within the interval $i\leq k\ll n$
which becomes relatively smaller as $n\to\infty$ and $\mu\to 0$.
Here $k\gg 1$ is the \emph{support of the bell-shaped function}.

The passage to the integral is accompanied by the appearance of a bell-shaped function that
progressively approaches the $\delta$-function (Bose-condensate).
Therefore in formula~\eqref{11} for $n>n_c$ or $T<1$
the values of $dn$ (and $dV$ in thermodynamics)
can be set equal to zero. As the result, the entropy of the ``liquid" phase
acquires the form:
\begin{equation}\label{12}
d(\mathcal{E}n)=T\, dS.
\end{equation}

On the plane $\{M, \, Z=\frac{M}{nT}\}$, the ``gas'' critical isotherm
\begin{equation}\label{12a}
M=\operatorname{Li}_{\gamma_c+2} (e^{-\mu/T_c}), \qquad
Z=\frac{\operatorname{Li}_{\gamma_c+2} (e^{-\mu/T_c})}{\operatorname{Li}_{\gamma_c+1} (e^{-\mu/T_c})}
\end{equation}
is practically the same for all pure gases each of which is associated with
$Z_c=\zeta(\gamma_c+2)/\zeta(\gamma_c+1)$.
For argon $Z_c=0.29$, for methane $Z_c=0.29$,  for oxygen and other gases,
the value of $Z_c$ is determined experimentally.
The ``base'' spinodal (i.e., the set of points for $\mu=0$)
on the plane $\{M, \, Z\}$ is the interval $Z=Z_c$, $M=\{0, 1\}$.
The isotherm $T\leq 1$ demonstrates a good coincidence on the interval
$-\infty<\mu<\mu_l$ up to the phase transition ``gas--liquid'' for $\mu=\mu_l$.

On the ``liquid'' isotherm, the value $n= T^{\gamma_c+1}\zeta(\gamma_c+1)$ is constant.
Therefore, it is a straight line passing through the point
$M=T^{\gamma_c+2}$, \,$Z=Z_c$ (the origin is $M=0, \, Z=0$).

\begin{remark}\label{rem2}
The Bose condensation in income statistics means that
when we increase the  number of trials $n>n_0$,
we necessarily come to the low income group
if the general population is the country's population,
while the order statistics is determined by the value of the random variable
(expressing the income of each citizen of the country; see Sec.~\ref{3.1}).
\end{remark}

\section{Negative values of the parameter $\gamma$}\label{new}

Under the assumption that the transition to the liquid phase
does not occur at $T=1$,
we equate the chemical potentials $\mu$ and $\tilde{\mu}$
for the ``liquid'' and ``gas'' phases on the isotherm $T=1$.

Then we determine the point $\mu$, i.e. the point of transition to the ``liquid'' phase
for $T< 1$, by equating the chemical potentials of the ``liquid'' and ``gas'' phases.

\begin{remark}
Consider the case $\gamma <0$ and $\tilde{\mu} =0$.

Let us introduce the notation
$$
 F(\xi)=\left(\frac 1 \xi -\frac{1}{e^{\xi}-1}\right).
$$

For $\gamma < 0$ and $\tilde{\mu} =0$ we have (according to \eqref{8j})
\begin{equation}
N_c= \Lambda^{\gamma-\gamma_c}\sum_{j=1}^\infty \frac{j^\gamma}{e^{bj}-1}=
\Lambda^{\gamma-\gamma_c} \left\{\sum_{j=1}^\infty j^\gamma\frac{1}{bj}-
\sum_{j=1}^\infty j^\gamma F(bj)\right\},\qquad b=\frac{1}{T}.
\end{equation}
Since the function $f(x)=x^\gamma F(bx)$ monotonically decreases, we have
\begin{equation}
\sum_{j=1}^\infty j^\gamma F(bj)=
\sum_{j=1}^\infty  f(j) \leq
\int_0^\infty f(x)\, dx=
\int_0^\infty x^\gamma F(bx)\, dx=
b^{-\gamma-1} \int_0^\infty x^\gamma F(x)\, dx
\end{equation}
(the Nazaikinsky inequality).

Thus, we have
\begin{equation}\label{26}
N|_{\tilde{\mu}=0}= \Lambda^{\gamma-\gamma_c} b^{-1}\zeta(1-\gamma)+O(b^{-1-\gamma}),
\qquad \gamma <0,
\qquad b=\frac 1T,
\end{equation}
where $\zeta$ is the Riemann zeta function.
\end{remark}

In this section we find the liquid isotherm-isobar point
as the quantity $\varkappa=-\mu/T$ slowly tends to zero.

First, we recall that although $N_c$ is large, it is still finite
and hence we must use the parastatistical correction.
We recall the relation
\begin{equation}\label{ad0}
\Omega=- \Lambda^{\gamma-\gamma_c}
T\sum_k\ln\Big(\frac{1-\exp\frac{\mu-\varepsilon_k}TN}{1-\exp\frac{\mu-\varepsilon_k}T}\Big)
\end{equation}
for the $\Omega$-potential.

We use the Euler--Maclaurin formula with the parameter~$\gamma$
taken into account and obtain
$$
\sum^{n}_{j=1}\Big(\frac{j^\gamma}{e^{bj+\varkappa}-1}
-\frac{kj^\gamma}{e^{bkj+k\varkappa}}\Big) =\frac1{\alpha}\int^\infty_0
\Big(\frac{1}{e^{bx+\varkappa}-1}-\frac{k}{e^{bkx+k\varkappa}-1}\Big)\,dx^\alpha+R,
$$
where $\alpha=\gamma+1$, $k=N$ and $b=1/T$.
Here the remainder $R$ satisfies the estimate
$$
|R|\leq \frac1\alpha\int^\infty_0|f'(x)|\,dx^\alpha, \qquad \text{where}\quad
f(x)=\frac1{e^{bx+\varkappa}-1}-\frac{k}{e^{k(bx+\varkappa)}-1}.
$$
We calculate the derivative and obtain
\begin{align}\label{ad1}
f'(x)&=\frac{bk^2e^{k(bx+\varkappa)}}{(e^{k(bx+\varkappa)}-1)^2}
-\frac{be^{bx+\varkappa}}{(e^{bx+\varkappa}-1)^2},
\nonumber\\[-3\jot]
\\
|R|&\leq \frac1{\alpha b^\alpha}
\int^\infty_0 \Big|\frac{k^2e^{k(y+\varkappa)}}{(e^{k(y+\varkappa)}-1)^2}
-\frac{e^{y+\varkappa}}{(e^{y+\varkappa}-1)^2}\Big|\,dy^\alpha.
\nonumber
\end{align}
We also have
$$
\frac{e^y}{(e^y-1)^2}=\frac1{y^2}+\psi(y),\qquad \text{where $\psi(y)$ is smooth and
$|\psi(y)|\leq C(1+|y|)^{-2}$}.
$$
Substituting this formula into~\eqref{ad1}, we obtain
\begin{align*}
|R|&\leq
\frac1{\alpha b^\alpha}
\int^\infty_0
\big| \psi\big(k(y+\varkappa)\big)-\psi(y+\varkappa)\big|\,dy^\alpha
\nonumber\\
&\leq
\frac{k^{-\alpha}}{b^\alpha}
\int^\infty_{k\varkappa} |\psi(y)|\,dy^\alpha
+\frac{1}{b^\alpha} \int^\infty_{\varkappa}|\psi(y)|\,dy \leq
Cb^{-\alpha}
\end{align*}
with a certain constant~$C$.
For example, if $\varkappa\sim(\ln k)^{-1/4}$,
then $|R|$ preserves the estimate $|R|\sim O(b^{-\alpha})$.
Therefore, we have $k=N_c$ and $T=T_c$
and obtain the following formula for the integral at $\mu=0$:
$$
M=
\frac {\Lambda^{\gamma-\gamma_c}}{\alpha\Gamma(\gamma+2)}\int\frac{\xi\,d\xi^\alpha}{e^{b\xi}-1}=
\frac {\Lambda^{\gamma-\gamma_c}}{b^{1+\alpha}}\int_0^\infty\frac{\eta d\eta^\alpha}{e^\eta-1},
$$
where $\alpha=\gamma+1$. Thus,
$$
b=\frac1{M^{1/(1+\alpha)}}\left(\frac {\Lambda^{\gamma-\gamma_c}}{\alpha\Gamma(\gamma+2)}
\int_0^\infty\frac{\xi\,d\xi^\alpha}{e^\xi-1}\right)^{1/(1+\alpha)}.
$$

We obtain (see~\cite{Arxiv_2009})
$$
\aligned
 &\int_0^\infty
\left\{\frac1{e^{b\xi}-1}-\frac{k}{e^{kb\xi}-1}
\right\}\,d\xi^\alpha =\frac1{b^\alpha}\int_0^\infty\left(\frac1{e^\xi-1}
-\frac1\xi\right)\,d\xi^\alpha \notag
\\
&\quad+\frac1{b^\alpha}\int_0^\infty\left(\frac1\xi-
\frac1{\xi(1+(k/2)\xi)}\right)\,d\xi^\alpha
-\frac{k^{1-\alpha}}{b^\alpha}\int_0^\infty\left\{
\frac{k^\alpha}{e^{k \xi}-1}
-\frac{k ^\alpha}{k \xi(1+(k /2)\xi)}\right\}\,d\xi^\alpha \notag
\\
=&\frac{c(\gamma)}{b^\alpha}(k^{1-\alpha}-1).
\endaligned
$$
Since $k\gg 1$, we finally have by setting $k=N|_{\tilde{\mu}/T=o(1)}$,
\begin{equation}\label{ad2}
N|_{\tilde{\mu}/T=o(1)}=(\Lambda^{\gamma-\gamma_c} c(\gamma))^{1/(1+\gamma)}T,
\quad\text{where}\quad
c(\gamma)=\int_0^\infty\Big(\frac1{\xi}-\frac1{e^\xi-1}\Big)
\xi^\gamma\,d\xi.
\end{equation}
The relation $N= T^{\gamma_c+1}\zeta(\gamma_c+1)$ agrees with the linear relation
$N=A(\gamma)T$, where $A(\gamma)= (\Lambda^{\gamma-\gamma_c} c(\gamma))^{1/(1+\gamma)}$,
for $P<0$.

We can normalize the activity $a$ at the point $T_c$ and can find $a_0$,
by adjusting the liquid and gas branches at $T_c$ for the pressure
(i.e., for $M$) so that the phase transition on the critical isotherm
could not occur at $T=1$.

Further, we normalize the activity for $T<1$ by the value of $a_0$ calculated below.
Then the chemical potentials
(the Gibbs thermodynamical potential for the liquid and gas branches in thermodynamics)
coincide, and hence no ``gas--liquid'' phase transition occurs at $T=1$
(cf. Sec.~5).

Now to realize the isochor--isotherm of ``incompressible liquid'',
we must also construct it a density lesser than $N_c=\zeta(\gamma_c+1)$, namely, for
$$
 N(T) =  T^{\gamma_c+1}\zeta(\gamma_c+1).
$$
We obtain the value $\gamma(T)$ from the implicit equation
$$
A(\gamma)= T^{\gamma_c}\zeta(\gamma_c+1).
$$

We thereby obtain the spinodal curve
(i.e., the points where $\tilde{\mu}\sim T (\ln N(T))^{-1/4}$)
for each $T<1$ in the region of negative $\gamma$:
\begin{equation}\label{1ad}
\Lambda^{-(\gamma-\gamma_c)/(1+\gamma)} c(\gamma)^{1/(1+\gamma)} =T^{\gamma_c}\zeta(\gamma_c+1).
\end{equation}

From the two values of $\gamma$ of the solution~\eqref{1ad}
we take the least (the largest in the absolute value) and denote it by~$\gamma (T)$.
In particular, for $T=1$ we denote $\gamma_0=\gamma(1)$.

The second solution can be realized for some metals (mercury, cesium, etc.)
and corresponds to the metastable state.
The actual interaction between particles gives a more significant deviation
of the second solution from~$\gamma_0$ (see~\cite{Shmul}).

Let $a_g=e^{-\mu/T}$ be the gas activity,
and let $a_l=e^{-\tilde{\mu}/T}$ be the liquid activity.
We give the condition that $M$ and the activities coincide
at the point of phase transition (see Sec.~5):
\begin{equation}\label{x}
T^{\gamma_c}\operatorname{Li}_{2+\gamma_c}(a_g)=
\Lambda^{-|\gamma(T)|-\gamma_c} T^{-|\gamma(T)|}\operatorname{Li}_{2-|\gamma(T)|} \left(\frac{a_l}{a_0} \right),
\end{equation}
\begin{equation}\label{xx}
\frac{\Lambda^{\gamma_0-\gamma_c}}{\zeta(2+\gamma_0)}
\operatorname{Li}_{2+\gamma_0}(a_0)=1,
\end{equation}
\begin{equation}\label{xxx}
a_g=\frac{a_l}{a_0}.
\end{equation}

These two equalities determine the value of the chemical potential
$\mu=\tilde{\mu}=T\ln a_g$ at which the ``gas--liquid'' phase transition occurs.

Let $T_0=\min_{-1<\gamma<0}A(\gamma)$.
Thus, for each $T_0<T<T_c$ we obtain the value of the liquid reduced activity
$a_r=a_l/a_0$ (where $a_l$ is the liquid activity),
which corresponds to the Van-der-Waals normalization
(see~\eqref{4ja}).

In thermodynamics, the critical values $T_c$, $P_c$, and $\rho_c$
were experimentally calculated almost for all gases,
and hence the critical number of degrees of freedom can be given in advance.
The parameter $\Lambda$ ($1.6 <\Lambda <3$, $T>1/3$) is obtained from
the coincidence of binodals at the triple point determined experimentally.

\begin{remark}
In the probability theory and economics,
these quantities can be determined only by indirect methods
(for example, by examining the income growth rate (see below)
or by an analysis of a sufficiently large set of debt-duration pairs
(duration is the period of debt repayment); see~\cite{MN_85-1}).
But the analogy with the Van-der-Waals law of corresponding states
allows us to hope that the scenarios of ``phase transitions''
must be sufficiently similar for different states
(see~\cite{Quant_Econ}).
\end{remark}

\section{Infinitely small dissipation taken into account and Maxwell rule}

A.S.Mishchenko and one of the authors~\cite{Mish}
considered the case of transition to the two-dimensional Lagrangian manifold
in the four-dimensional phase space,
where the pressure~$P$ and the temperature~$T$ (intensive variables)
play the role of coordinates
and the extensive variables, the volume~$V$ and the entropy~$S$
play the role of momenta of the Lagrangian manifold,
where the entropy is the action generating the Lagrangian structure.

It turns out that this complicated transformation,
which leads to the Gibbs thermodynamical potential,
can be performed only by using the extension
to the region of negative values of~$\gamma$.
Then one can justify the Maxwell transition by introducing a small dissipation
(viscosity).
The introduction of infinitely small dissipation
allows one simultaneously to solve the problems of critical exponents
without using the scaling hypothesis, which underlies the renormalization group method.
Let us show this.

There is no viscosity in thermodynamics.
However, in general, thermodynamical equilibrium
cannot occur without infinitely small dissipation.
Therefore, it is natural to introduce a small viscosity
and then to let it tend to zero.

The geometric quantization of the Lagrangian manifold
(see~\cite{Hart}, \S 11.4) is usually associated with the introduction
of the constant~$\hbar$.
The case of pure imaginary number~$\hbar$
was called the Wiener (or tunnel) quantization
by one of the authors~\cite{Asimp_Met}.

Let us apply the Wiener quantization to thermodynamics.
The thermodynamical potential $G=\mu N$ is the action
$\mathbb{S}=\int p\,dq$ on the two-dimensional Lagrangian manifold~$\Lambda^2$
in the four-dimensional phase space $q_1$, $q_2$, $p_1$, $p_2$,
where $q_1$ and $q_2$ are the pressure~$P$ and the temperature~$T$, respectively,
$p_1$ is equal to the volume $V$, and~$p_2$ is equal to the entropy~$S$
with opposite sign.
All other potentials, i.e., the internal energy~$E$,
the free energy~$F$, and the enthalpy~$W$ are the results of projection
of the Lagrangian manifolds on the coordinate planes~$p_1$,~$p_2$
\begin{gather}\label{pr}
E=-\int \vec{q}\, d\vec{p}, \quad
\vec{q}=\{q_1, q_2\}, \vec{p}=\{p_1, p_2\},
\notag \\
W=-\int(q_2\, dp_2+ q_1\, dp_1), \qquad
F= \int (q_1\, dp_1 - q_2\, dp_2).
\end{gather}

In the Wiener quantization, we have
$$
N=\varepsilon\frac{\partial}{\partial \mu}, \qquad
V=\varepsilon\frac{\partial}{\partial p}, \qquad
S=-\varepsilon\frac{\partial}{\partial T}.
$$
Hence the role of time~$t$ is played by $\ln(-\mu/T)$
$$
G = \mu N\sim \varepsilon\frac{\mu}{T}\frac{\partial}{\partial (\mu/T)}
=\varepsilon\frac{\partial}{\partial \ln(\mu/T)}.
$$

Note that the tunnel quantization of the Van-der-Waals equation (VdW)
gives the Maxwell rule as $\varepsilon \to 0$ (see below).

As is seen below, the critical point and spinodal points are focal point
and hence do not ``pass'' into classics, i.e., into the VdW model,  as
$\varepsilon \to 0$. The spinodal points, similarly to the turning points
in quantum mechanics, are approximated by the Airy function,
and the critical point, which is the point of origination of two turning points
(two Airy functions) is approximated by the Weber function (see~\cite{UMN}).
The point of shock wave origination in the B\"urgers equation
is expressed precisely in terms of the Weber function
as $\varepsilon \to 0$.
If the transition as $\varepsilon \to 0$ is performed outside these points,
then the VdW--Maxwell model is obtained.
But the limit transition is violated at these points themselves.
Therefore, the so-called ``classical'' critical Landau exponents~\cite{Landau_St-ph}
turn out to disagree seriously with experiments.
The Weber function gives singularities of the form
$\varepsilon^{-1/4}$, and the Airy function, of the form~$\varepsilon^{-1/6}$.

We now consider the B\"urgers equations in more detail.

Consider the heat equation
\begin{equation}\label{opm1}
\frac{\partial u}{\partial t}=\frac \varepsilon 2\frac{\partial^2 u}{\partial x^2}, \qquad
 x\in \mathbb{R}, \quad t \geq 0,
\end{equation}
where $\varepsilon>0$ is a small parameter. It is known that all linear combinations
\begin{equation}\label{opm2}
u=\lambda_1 u_1 + \lambda_2 u_2
\end{equation}
of the solutions $u_1$ and $u_2$ to Eq.~\eqref{opm1}
are solutions of this equation.

We perform the change
\begin{equation}\label{opm3}
u=\exp (-w(x,t)/\varepsilon)
\end{equation}
and obtain the nonlinear equation
\begin{equation}\label{opm4}
\frac{\partial w}{\partial t} +
\frac 12 \bigg(\frac{\partial w}{\partial x}\bigg)^2-
\frac \varepsilon2 \frac{\partial^2 w}{\partial x^2} =0,
\end{equation}
which is called the integrated B\"urgers equation
\footnote{The usual B\"urgers can be derived from Eq.~\eqref{opm4}
by differentiating with respect to~$x$ and performing the change
$v =\partial w/\partial x$.}.
Obviously, any solution $u_i$ of Eq.~\eqref{opm1} is associated with the solution
$w_i=-\varepsilon \ln u_i$  of Eq.~\eqref{opm4},  $i=1,2$.
The solution~\eqref{opm2} of Eq.~\eqref{opm1}
is associated with the solution of Eq.~\eqref{opm4}
$$
w= -\varepsilon \ln\big(e^{-\frac{w_1+\mu_1}{\varepsilon}} + e^{-\frac{w_2+\mu_2}{\varepsilon}}\big),
$$
where $\mu_i= -\varepsilon \ln \lambda_i$, ($i=1.2$). Since
$$
\lim_{\varepsilon\to 0}w =\min (w_1, w_2),
$$
we obtain the $(\min, +)$ algebra of tropical mathematics~\cite{Litvinov}.

To determine the solutions of $t>t_{\mathrm{cr}}$,
Hopf proposed to consider the B\"urgers equation
\begin{equation}\label{burg}
\frac{\partial v}{\partial t} + v\frac{\partial v}{\partial x}- \frac \varepsilon 2
\frac{\partial^2 v}{\partial x^2}=0, \qquad v|_{t=0} =p_0(x).
\end{equation}
The (generalized) solution of the equation
\begin{equation}\label{burg_1}
\frac{\partial p}{\partial t} + p\frac{\partial p}{\partial x}=0,
\qquad p|_{t=0} =p_0(x)
\end{equation}
is the function $p_{\text{gen}}=\lim_{\varepsilon\to 0} v$ (the Riemann waves).

The solution $v$ of the B\"urgers equation is expressed
in terms of the logarithmic derivative
\begin{equation}\label{burg_2}
v=-\varepsilon\frac{\partial}{\partial x}\ln u
\end{equation}
of the solution $u$ of the heat equation
\begin{equation}\label{teplo}
\frac{\partial u}{\partial t} =\frac \varepsilon 2 \frac{\partial^2 u}{\partial x^2},
\qquad u|_{t=0} = \exp \bigg\{- \frac 1 \varepsilon \int_{-\infty}^x p_0(x)\, dx\bigg\}.
\end{equation}
Thus, the initial problem is reduced to studying the logarithmic limit
of the solution of the heat equation.
It is known that the solution of problem~\eqref{teplo} has the form
\begin{equation}\label{teplo_1}
u=(2\pi \varepsilon t)^{-1/2} \int_{-\infty}^{\infty}
\exp \bigg\{ - \bigg( (x-\xi)^2+2t \int_{-\infty}^{\xi} p_0(\xi)\, d\xi\bigg) \bigg/ 2th\bigg\}\, d\xi.
\end{equation}
The asymptotic behavior of the integral~\eqref{teplo_1} is calculated by the Laplace method.
For $t<t_{\mathrm{cr}}$, we have
\begin{equation}\label{teplo_2}
u= \big(|J|^{-1/2} (\xi(x,t),t) + O(\varepsilon)\big)
\exp \bigg\{ -\frac 1\varepsilon S(x,t)\bigg\}.
\end{equation}

Here $S(x,t)= \int_{-\infty}^{r(t)} p\, dx$,
and the integral is calculated along the lagrangian curve $\Lambda^t$,
and the point $r(x)$ is a point on $\Lambda^t$.
For $t>t_{\mathrm{cr}}$, there are three points
$r_1(x)$, $r_2(x)$, $r_3(x)$ on $\Lambda^t$ whose projections on the axis~$x$
coincide or, in other words, the equation $Q(t,\xi)=x$ for $x\in (x_1, x_2)$
has three solutions $\xi_1(x,t)$, $\xi_2(x,t)$, $\xi_3(x,t)$.

We introduce the notation $S(x,t)=\int_{-\infty}^{r(x)} p\, dx$ for $x<x_1$, $x>x_2$,
and $S(x,t)= \min(S_1, S_2, S_3)$,  $S_j=\int_{-\infty}^{r_j(x)} p\, dx$, where
$j=1,2,3$ for $x \in[x_1, x_2]$.

The above arguments allow us to obtain a generalized discontinuous solution of
problem~\eqref{burg_1} for times $t>t_{\mathrm{cr}}$.
It is given by the function $p=p(x,t)$
determining essential domains~\cite{Asimp_Met} of the curve~$\Lambda^t$.
We note that, in particular, this implies the rule of equal areas
known in hydrodynamics to define the front of the shock wave
whose evolution is described by Eq.~\eqref{burg_1}.
We note that this exactly corresponds to the Maxwell rule
for the VdW equation.

The solution $v=v(x,\varepsilon)$ of the B\"urgers equation at the critical point
$x=p^3$ is calculated by the formula
\begin{equation}\label{m15}
v(x,\varepsilon)=\varepsilon\frac{\partial\ln u(x)}{\partial x}
= \frac{\int_0^\infty\exp \{\frac{-x\xi -\xi^4/4}{\varepsilon}  \}\xi\, d\xi}
{\int_0^\infty\exp \{\frac{-x\xi -\xi^4/4}{\varepsilon} \}d\xi}.
\end{equation}
As $x\to 0$, after the change $\frac{\xi}{\sqrt[4]{\varepsilon}}=\eta$,
we obtain
\begin{equation}\label{m16}
v(\varepsilon,x) \rightarrow_{x\to 0}  {\sqrt[4]{\varepsilon}}\cdot \text{const}.
\end{equation}

What does this mean from the standpoint of classics and classical measurements,
i.e., in the case where the condition
called a condition for being ``semiclassical'' in the book~\cite{Landau_Qu_Mec}
is satisfied
(i.e., in the case where we are outside the focal point)?
For the Laplace transformation this means that we are in the domain
where the asymptotic Laplace method can be applied, i.e., where
\begin{equation}\label{gg1}
u(x)= \frac {1}{\sqrt{\varepsilon}}
\int_0^\infty  e^{-\frac{px-\widetilde{S}(p)}{\varepsilon}} \, dp.
\end{equation}

Moreover, the solution of the relation
\begin{equation}\label{m16a}
x=\frac{\partial \widetilde{S}}{\partial p}
\end{equation}
is not degenerate, i.e.,
$\frac{\partial^2 \widetilde{S}}{\partial p^2}\neq 0$ at the point
$\frac{\partial\widetilde{S}}{\partial p}=x$,
and the reduced integral~\eqref{gg1} is bounded as $\varepsilon\to 0$.
For it to have a zero of the order of $\varepsilon^{1/4}$,
we must integrate is over~$x$ with the fractional derivative $D^{-1/4}$.
The value of $D^{-1/4}$ applied to unity, $D^{-1/4}1$,
is approximately equal to $x^{1/4}$.

According to the indeterminancy principle~\cite{TMF_170-3},
the correspondence between the differentiation operator and
a small parameter of the form $D \to 1/\varepsilon$
is preserved for the ratio $-\varepsilon \frac{\partial u/\partial x}{u}$,
although the leading term of the asymptotic expansion in the difference of
$\frac{\partial^2 u/\partial x^2}{u}$ and $\frac{(\partial u/\partial x)^2}{u^2}$
is eliminated.

In the case of thermodynamics, the role of $x$ is played by the pressure $P$
and the role of the momentum $p$ is played by the volume~$V$.
Therefore, $V\sim P^{1/4}$, i.e.,
\begin{equation}\label{m16b}
P_{c} \sim (V -V_{c})^4.
\end{equation}

This is precisely the jump of the critical exponent.
The other critical exponents can be obtained similarly~\cite{TMF_170-3}.
The comparison with experimental data is discussed in the same paper.

\section{Independent events, conditional probability, and logarithmic scaling}\label{Ind_Ev}

\begin{definition}
We will call two events {\it independent}
if the Hartley entropies corresponding to the events add together.
\end{definition}

Let us consider the conditional probability for independent events.

It is well known that the {\it conditional probability} of an event
under a condition is the probability of the simultaneous occurrence
of the event and the condition normed by the probability of the condition:
$$
P(A|B) = \frac{P(A\bigcap B)}{P(B)},
$$
where $A$ is the event and $B$ is the condition.

The formula
\begin{equation}\label{17}
P(B)= \sum_{i=1}^k P(B\bigcap A_i)=\sum_{i=1}^k P(B|A_i) P(A_i), \qquad  \prod A_i=\Omega
\end{equation}
for nonintersecting events $\{A_i\}$ is called the {\it total probability} formula.

A similar formula holds in the new conception of probability. It suffices to prove it for $k=2$.
This, according to Kolmogorov's complexity theory, must be the equality of entropies with
conditional probability $\{\alpha,\,\beta=1-\alpha\}$ taken into account.
Namely, for $\tilde{\mu}=0$ we have
\begin{equation}\label{18}
Z_{\mathrm{c}}(\gamma_{\mathrm{c}}+2)=
\alpha(\gamma^{\mathrm{c}}_1+2)Z^{\mathrm{c}}_1+
\beta(\gamma^{\mathrm{c}}_2+2)Z^{\mathrm{c}}_2.
\end{equation}
where $Z_{\mathrm{c}}=\zeta(\gamma+2)/\zeta(\gamma+1)$, $\alpha+\beta=1$.

From these formulas, we obtain the value of
$\gamma_{\mathrm{c}}$ for ``mixtures" such as those given by~\eqref{17}
by using the basis variational series.

The above construction can easily be generalized to the general case
of formula~\eqref{17}.

Logarithmic scaling determines, in the order statistics (variational series),
a bijection between the numbers $A\geq 0$ and $\ln A=a$, $B\geq 0\to b=\ln B$.
To the product $AB$ corresponds, under logarithmic scaling, the sum $\ln A + \ln B$.

\textbf{Addition under logarithmic scaling.}

The summing algorithm is given by
\begin{equation}\label{sum1}
A\oplus B = \frac{1}{\ln n} \ln(n_c^{(1)} + n_c^{(2)})
\end{equation}
with the restriction~\eqref{18} on the growth of~$n$ taken into account.

By means of this bijection, we can encode any numbers $A$ and $B$,
as well as the semifield $AB$ and the correspondence $A+B \to A\oplus B$.

The addition~\eqref{sum1} as $n\to\infty$
undergoes a ``phase jump" from the ``gas" state to the ``liquid" state,
i.e., from an ``open state" to a ``closed" one.

\begin{remark}\label{Plot}
Note that the volume $V$, one of the main notions in thermodynamics, is absent
from our considerations, and the role of density is played by the number of particles $N$. During
the phase transition gas-liquid, the density undergoes a jump and therefore, in our case, so does
the quantity $N$. In thermodynamics, it is natural to assume that the number of particles is
constant, and the jump in density is due to a jump in the volume $V$. Then the chemical potential
$\mu$, multiplied by the number of particles, is equal to the Gibbs thermodynamical potential, which
remains continuous at the point of phase transition (the Maxwell rule).
Therefore, the chemical potential of the liquid and the gas phases must coincide.
\end{remark}

\begin{remark}
This entropy, according to Kolmogorov's complexity theory, is intimately linked
with encodings of minimal length.
Thus the phase transition of the ``gas'' entropy~\eqref{11}
into the ``liquid'' entropy~\eqref{12} is the transition
to a new encoding. We cannot decrease the number of trials:
{\it the number of trials is irreversible.}
In our case an increase in the number of trials
in the state of ``liquid phase'' gives us nothing, because the
extra trials fall out into the ``Bose condensate" of the ``Bose liquid''.
\end{remark}

\begin{remark}
In thermodynamics, the volume $V$ is also introduced,
but the geometry of the volume $V$ is neglected in principle,
i.e., the problem is invariant under the variations of its shape.
There is no external field in this problem, and hence the one-particle
potential in the Gibbs distribution must have the form $\varphi(r^3/V)$
if the molecule radius is ``infinitely small''.
And since the volume is large, the leading term as $V\to \infty$ has the form
\begin{equation}\label{92-5-8}
-\frac{\alpha r^3}{V},
\end{equation}
where $\alpha$ is a parameter which can be expressed in terms of a new parameter $q$.
The parameter $q$ is called the heat of transition from
the gas phase into the liquid phase~\cite{Landau_St-ph} as $T\to 0$.
Since the additional parameter must be independent of the shape of the vessel,
it suffices to take the volume of a ball of radius~$R$
for determining this parameter.

Taking account of this potential changes
the pressure and the compressibility factor~$Z$,
which allows one to take account of the reflection from the vessel walls
in the general form. We have
\begin{equation}\label{v1}
\int_0^{R}  e^{-r^3\xi k/R}\, dr^3=
\frac{1}{k(\beta-\beta_c)} \int_0^{k(\beta-\beta_c)} e^{-y}\, dy =
\frac{1}{k(\beta-\beta_c)}\left(e^{(\beta_c-\beta)k}-1\right),
\end{equation}
where $\xi=q(\beta-\beta_c)$,  $\beta=1/T$,  $\beta_c=1/T_c$.

Hence the $k$th term in the definition of polylogarithm
$$
\operatorname{Li}_s(a)= \sum_{k=1}^\infty \frac{a^k}{k^s}
$$
becomes
$$
\frac 1\xi \cdot \frac{(e^{\xi k}-1)a^k}{k^{3+\gamma_c}},
$$
whence
\begin{align}\label{v2}
P&=-\frac{T^{2+\gamma_c}}{\xi} \left\{\operatorname{Li}_{3+\gamma_c}(ae^{-\xi})
-\operatorname{Li}_{3+\gamma_c}(a)\right\},
\\
\label{v3}
Z&=\frac{\operatorname{Li}_{3+\gamma_c}(ae^{-\xi})
-\operatorname{Li}_{3+\gamma_c}(a)}{\operatorname{Li}_{2+\gamma_c}(ae^{-\xi})
-\operatorname{Li}_{2+\gamma_c}(a)}.
\end{align}
Now the condiiton for the spinodal in the domain $\gamma < 0$ becomes
\begin{equation}\label{v4}
A(\gamma)= T^{\gamma_c}\frac{|\operatorname{Li}_{2+\gamma_c}(e^{-\xi})
-\operatorname{Li}_{2+\gamma_c}(1)|}{\xi}.
\end{equation}
\end{remark}

\section{Applications}

\subsection{Distribution according to incomes}\label{3.1}

One of the most important problems of contemporary mathematical
statistics is that the ``general population" of measurable random variables
is very large, and the samples are also large, but much
smaller than the general population.

In the USSR the sample used in social questions was usually 3.5 thousand people.

As an example of the general population, one could consider the
set of all citizens of the Russian Federation. We would like to
calculate their mean income.

There is a quantity fixed by the state: the minimal wage and the
nominal percentage~$\tilde{\mu}$.

On the other hand, the income of many citizens is less than the
minimum wage, e.g., the family members of the worker (dependents),
as well as the unemployed. If we construct the order series,
the ``order statistics"  according to increasing incomes\footnote{This
can actually be done only in the countries where the incomes can be calculated
from the paid taxes, which means that the incomes are not hidden as a rule.}
and slit it up into clusters of 1--3\%, then, denoting the values of the
random variable of $\xi$-income by  $\xi_i$, we obtain the variational series
$$
\bigg\{\frac{\xi_i}{T}\bigg\}^\infty_{i=0}.
$$
Suppose that the number of trials (size of the sample) is~$n$,
while $n_i$ is the number of outcomes for which the number
${\xi_i}/{T}$ is obtained; then, according to the usual
definition, the estimate of the expectation of the sample,
multiplied by $n$ trials, is equal to
\begin{equation}\label{1}
\sum n_i \frac{\xi_i}{T} =M, \qquad \sum n_i =n.
\end{equation}
Thus the estimate of the expectation of the sample is
\begin{equation}\label{2}
E=\frac{M}{n}.
\end{equation}

Obviously, if the $\xi_i$ take unboundedly large values, then a
randomly chosen sample cannot serve even as an approximate
``estimate" of the correct expectation.

Thus, in determining the average per capita income, we must divide
the general population into parts, distinguishing the region that
characterizes (in the mean) the studied stratum of society, for
which we accept the hypothesis of homogeneous distribution.
Sometimes one considers separate households in their development
(longitudinal studies).

Sometimes one also uses the following methodology for measuring
the distribution of the population according to income. Here the
initial data are regular reports, not supported by any documents,
of 50 000 people, of which 2000 are muscovites. The choice of
these respondents is purposely left unchanged (new ones are added
only when some of the previous ones disappear). There aren't any
very rich people among them, because participation is purely
voluntary. There may be some people among them who live below the
poverty line, because the reports are modestly paid for.

The obtained data is extrapolated by comparing the socio-economic
characteristics of the respondents (region and type of city town
or village, gender, age, size of the family, profession, etc.) are
compared with the data given by census.

No estimates of the precision of such studies are made, either in
Russia or abroad. There is only the comparison with
``theoretically expected" distributions (lognormal with a Pareto tail).
Pareto was the first to propose the power distribution as
the description of the distribution of incomes in accordance with
empirical data. There have also been theoretical discussions about
the possibility of interpreting the observed discrepancies by  the
presence of numerous maxima.

However, it is impossible to partition an infinite set
into a finite family of finite sets.

Therefore, in our case it is natural to call the estimate of the expectation
the empirical expectation.

We have called the quantity
\begin{equation}\label{3}
\lim_{n\to\infty} \frac{\ln M}{\ln n} =D
\end{equation}
the {\it dimension} in the situation corresponding to quantum mechanics and the
{\it number of degrees of freedom} in the statistical case.

It turns out that for any $M$ (under our assumption~\eqref{3}
there exists a critical number of measurements $N_{\mathrm{c}}\ll M$
such that, when the number of measurements increases above this critical value
$N>N_{\mathrm{c}}$ with our asymptotic precision
(e.g., $O(N^{\alpha})$, where $\alpha<1$), we will reach the level of the unemployed
or very low income groups, which within the limits of our accuracy does not influence the value
of~$N_{\mathrm{c}}$.

\begin{remark}
Let us explain what we mean by a closed system. This is not the analog of the
situation the ruble in the USSR, although the ruble was not convertible. The USSR did not have a
market economy, in other words, it did not have an economic system based on entropy.

Another type of closed system is the closed system of currency in postwar Great Britain. At that
time the exchange of the pound for other currencies was forbidden. Similarly, in Israel, the
exchange of the shekel  was only allowed within the limit of\$\,500.

A rough analogy with liquid and gas may be carried out here. If we open the cork of a bottle of gas,
the gas will leave the bottle, unlike the contents of a bottle of water. The surface tension, which
is the boundary of the liquid, is similar to the boundary of a country with a closed system of
currency.
\end{remark}

\subsection{Problems of queuing theory on the Web}

Here for the general population we take all the communicators: computers, i-pads,
i-phones, smart cell phones, etc.

When there  is only  one communication channel, then a message sent before the others will arrive
first. Here the order is preserved, and the Shannon theory is applicable, in particular we have the
Shannon entropy, which is equal to the Boltzman entropy. But if the messages are gathered in
packets, then this principle can no longer be applied, and we come to a specific situation. In
problems of queuing theory, the value of the random variable defined as the number of dispatched
packets, corresponding to the given general population, grows without bound. We have in mind the
order statistics and the corresponding variational series (see~\cite{MN_91-5}).

A supercomputer connected to the World Wide Web is similar to a member state in a world wide union
of states. Sufficient conditions for the computer's dropping out of the Web follow from the
unrestricted probability theory described above.

\begin{remark}\label{rem6}
Under the total reign of the World Wide Web,
when even high level functionaries must confess to it under
control of its lie detector, ethical rules prescribed by various
religions as ``rules of the game" and ensuring the existence of
human society, can be monitored via the Web.
\end{remark}

\subsection{Relationship with the Monte-Carlo method}

As an example of the relationship between the Monte-Carlo method,
let us describe a specific problem for a self-teaching
audio-guide, a device that describes paintings in art museums.
This device is oriented to three degrees of freedom: motion in the
direction ``North--South", motion in the direction ``East-West"
and rotation about its axis. However, the device must also take
into account the directions along which  there are no paintings
(e.g., the directions of certain corridors).

On the one hand, three degrees of freedom give values of the
random variable that grow like  $\{ i^{3/2}\}$. On the other hand,
the monitoring of corridors without paintings is naturally ensured
by a separate program in its self-teaching software.

The application of the Monte-Carlo method for choosing a
pseudorandom variable in the computer language Java preserves the
number of degrees of freedom. For a sufficiently large number of
trials (i.e., visitors to the museum), the device's software,
supposedly adapted to the picture gallery, can become, as the
result of self-teaching, a software adapted to the architecture of
that specific museum. And this is the passage from an open system
to a closed one.

If the usual rounding procedure is used, i.e.,
if pseudorandom variables of some finite length are considered,
then the testing of the initial program is not sufficiently perfect,
and hence the device cannot meet all requirements of the consumer.
In the example of an art museum, this fact is not catastrophic.
Only the reputation of the firm producing this device will suffer.
But if such a device is used as a guide of a blind man,
then such a rounding imperfection can lead to a catastrophe.

\bigskip
The authors express their deep gratitude to Academician A.A.Guseinov,
professors G.I.Arkhipov, V.S.Vorobiev, V.N.Chubarikov, and M.S.Kosolapov
for fruitful discussions.
We also especially express our thanks to the editor T.Tolozova
for valuable remarks about the composition of our humanities-mathematical paper.

This work is supported by the Russian Foundation
for Basic Research (grant no.~11-01-12058\_\text{ofi}\_\text{m}).


\begin{thebibliography}{99}


\bibitem{Poincare}
H.~Poincar\'e,
{\it On Science}
(Nauka, Moscow, 1983)
[in Russian]
(Russian translation of his books
{\it Science et Hypoth\`ese, Valeur de la Science, Science et M\'ethode,
Derni\`eres Pens\'ees}).

\bibitem{Gusein}
A.~A.~Guseinov,
{\it Great Prophets and Thinkers, Ethical Studies from Moses to the Present Day}
(Moscow, Veche Editions, 2009)
[in Russian].

\bibitem {Kozlov_2009}
V.~V.~Kozlov,
``Kinetics of Collisionless Gas: Equalization of Temperature, Growth of the
Coarse-Grained Entropy and the Gibbs Paradox,''
Regul. Chaotic Dyn. \textbf{14} (4--5), 535--540  (2009).

\bibitem{MZ_83-5}
V.~P.~Maslov,
``Solution of the Gibbs Paradox in the Framework of Classical Mechanics
(Statistical Physics) and Crystallization of the Gas $C_{60}$,''
Mat. Zametki {\bf83} (5), 787--791 (2008)
[Math. Notes {\bf83} (5--6), 716--722 (2008)].

\bibitem{RJ-17-3}
V.P.Maslov,
``Solution of the Gibbs Paradox Using the Notion of Entropy
as a Function of Fractial Dimension'',
Russian J. Math. Phys. \textbf{17} (3), 251--261 (2010).

\bibitem{MZ_89_3}
V.~P.~Maslov,
``Gibbs Paradox, Liquid Phase as an Alternative To The Bose Condensate,
And Homogeneous Mixtures Of new ideal gases'',
Math. Notes, \textbf{89}, (3), 366--373 (2011).

\bibitem{TMF_170-3}
V.~P.~Maslov,
``Critical Exponents as the Wiener Quantization of Thermodynamics,''
Teoret. Mat. Fiz. {\bf170} (3), 458--470 (2012)
[Theoret. and Math. Phys. {\bf170} (3), 384--393 (2012)].

\bibitem{DAN_1957}
V.P. Maslov,
``Degeneration in the Transition from the Discrete Spectrum
to the Continuous One and the Transition from Quantum Mechanics to Classical Mechanics,"
Dokl. Akad. Nauk SSSR  {\bf114} (5), 957--960 (1957)
[Soviet Math. Dokl.].

\bibitem{NTI}
T.~V.~Maslova,
``A Refinement of the Zipf Law for Frequency Dictionaries,"
NTI  Ser.~2 {\bf 37} (11), (2006)
[Scientific and Technical Information Processing (11),(2006)].

\bibitem{Erdos-Leh}
P.~Erd{\H o}s and J.~Lehner,
``The Distribution of the  Number of Summands in the
Partitions of a Positive Integer,''
Duke Math. J. \textbf{8} (2), 335--345 (June 1941).

\bibitem{MN_91-5}
V.~P.~Maslov,
``Unbounded Probability Theory Compatible with the Probability Theory of Numbers,''
Math. Notes \textbf{91} (5), 603--609, (2012).

\bibitem{Landau_St-ph}
L.~D.~Landau, and E.~M.~Lifshits,
{\it Statistical Physics}
(Nauka, Moscow, 1964)
[in Russian].

\bibitem{Vershik}
A.~M.~Vershik,
``Statistical Mechanics of Combinatorial Partitions, and Their Limit Shapes,''
Funktsional. Anal. Prilozhen. \textbf{30} (2), 19--39 (1996)
[Functional Anal. Appl. \textbf{30} (2), 90--105 (1996)].

\bibitem{Masl_Naz_83-2}
V.~P.~Maslov and V.~E.~Nazaikinskii,
``On the Distribution of Integer Random Variables Related by a Certain Linear Inequality: I,''
Mat. Zametki \textbf{83} (2) 232--263 (2008)
[Math. Notes \textbf{83} (2) 211--237 (2008)].

\bibitem{RJ_19-3-VM}
V.~P.~Maslov,
``Bose Condesate in the $D$-Dimensional Case, in Particular, for D = 2,''
Russ. J. Math. Phys. \textbf{19} (3), 1--10 (2012).

\bibitem{Shmul}
K.~I.~Shmulovich and L.~Mercury,
``Geochemical Processes at Negative Pressures,''
Electronic Scientific Information Journal
``Herald of the Departure of Earth Sciences RAS'' {\bf1} (24), 1--3 (2006).

\bibitem{Arxiv_2009}
V.~P.~Maslov,
\textit{Threshold Levels in Economics},
\texttt{\tt arXiv:0903.4783v2 [q-fin.ST], 3~Apr 2009}.

\bibitem{MN_85-1}
V.~P.~Maslov,
``Theorems on the Debt Crisis and the Occurrence of Inflation,''
Math. Notes \textbf{85} (1), 146--150 (2009).

\bibitem{Quant_Econ}
V.~P.~Maslov,
\textit{Quantum Economics}
(Nauka, Moscow, 2006).

\bibitem{Mish}
V.~P.~Maslov and A.~S.~Mischenko,
``Geometry of a Lagrangian Manifold in Thermodynamics,''
Russ. J. Math. Phys. \textbf{10} (2), 161--172 (2003).

\bibitem{Hart}
N.~Hurt,
\textit{Geometric Quantization in Action}
(Reidel, Dordrecht, 1983; Mir, Moscow, 1985).

\bibitem{Asimp_Met}
V.~P.~Maslov,
\textit{Asymptotic Methods and Perturbation Theory}
(Nauka, Moscow, 1988)
[in Russian].

\bibitem{UMN}
V.~P.~Maslov,
``Nonstandard Characteristics in Asymptotic Problems,''
Uspekhi Mat. Nauk \textbf{38}(6), 3--36 (1983)
[Russian Math. Surveys \textbf{38}(6), 1--42 (1983)].

\bibitem{Litvinov}
G.~L.~Litvinov,
 ``The Maslov Dequantization, Idempotent and Tropical Mathematics:
A Very Brief Introduction,''
in \textit{Contemp. Math.}, Vol.~377:
\textit{Idempotent Mathematics and Mathematical Physics}
(Amer. Math. Soc., Providence, RI, 2005).

\bibitem{Landau_Qu_Mec}
L.~D.~Landau, and E.~M.~Lifshits,
{\it Quantum Mechanics}
(Nauka, Moscow, 1976)
[in Russian].

\end{thebibliography}
\end{document}